\documentclass[10pt,twocolumn,twoside]{IEEEtran}
\usepackage{amsmath,graphicx}
\usepackage{epstopdf}
\usepackage{multirow}
\usepackage{balance}
\usepackage{algorithm}
\usepackage{cite}
\usepackage[caption=false,font=footnotesize]{subfig}
\usepackage{stfloats}
\usepackage{amssymb}
\usepackage{algorithmicx}
\usepackage{algpseudocode}
\usepackage{amsfonts}
\usepackage{marvosym}
\graphicspath{{figure/}}

\usepackage{xcolor}

\begin{document}
	
	\title{Rank Minimization-based Toeplitz Reconstruction for DoA Estimation Using Coprime Array}
	
	\author{Shengheng~Liu,~\IEEEmembership{Member,~IEEE},
        Zihuan~Mao,~\IEEEmembership{Student Member,~IEEE},\\
		Yimin~D.\ Zhang,~\IEEEmembership{Fellow,~IEEE},
		Yongming~Huang,~\IEEEmembership{Senior Member,~IEEE}
		\vspace{-1.5em}
		
		\thanks{This work was supported in part by the National Natural Science Foundation of China under Grant Nos. 62001103, U1936201 and the Basic Research Program of Jiangsu Province under Grant No. BK20190338. (Corresponding author: Yongming Huang).}
		\thanks{Z.~Mao, S.~Liu, and Y.~Huang are with the School of Information Science and Engineering, Southeast University, Nanjing 210096, China, and the Purple Mountain Laboratories, Nanjing 211111, China (emails: \{mzh; s.liu; huangym\}@seu.edu.cn).}
		\thanks{Y.~D.\ Zhang is with the Department of Electrical and Computer Engineering, College of Engineering, Temple University, Philadelphia, PA 19122 USA  (email: ydzhang@temple.edu).}
	}
	
	\markboth{submitted to IEEE Communications Letters,~Vol.~25, No.~XX, XXX~2021}%
	{Liu \MakeLowercase{\textit{et al.}}: Rank Minimization-based Toeplitz Reconstruction for DoA Estimation Using Coprime Array}
	
	\maketitle
	
\begin{abstract}		
In this paper, we address the problem of direction finding using coprime array, which is one of the most preferred sparse array configurations. Motivated by the fact that non-uniform element spacing hinders full utilization of the underlying information in the receive signals, we propose a direction-of-arrival (DoA) estimation algorithm based on low-rank reconstruction of the Toeplitz covariance matrix. The atomic-norm representation of the measurements from the interpolated virtual array is considered, and the equivalent dual-variable rank minimization problem is formulated and solved using a cyclic optimization approach. The recovered covariance matrix enables the application of  conventional subspace-based spectral estimation algorithms, such as MUSIC, to achieve enhanced DoA estimation performance. The estimation performance of the proposed approach, in terms of the degrees-of-freedom and spatial resolution, is examined. We also show the superiority of the proposed method over the competitive approaches in the root-mean-square error sense.	
\end{abstract}
\vspace{-0.5em}
\vspace{-0.5em}
	
\section{Introduction}
\label{sec:intro}
	
\IEEEPARstart {D}{irection}-of-arrival (DoA) estimation is recognized as an important and fundamental problem in array signal processing with many other engineering applications. Aiming at detecting more targets than the number of sensors, sparse arrays are employed in the context of difference coarray\cite{Vaidyanathan2011Sparse,2013Sparsity,Qin15,2017Source,2017Sparsity,2018DOA}. Coprime array is considered as a preferred choice because it offers reduced mutual coupling and provable performance guarantees as well as the merits that are shared by other sparse arrays, such as enlarged array aperture and increased degrees-of-freedom (DoFs) compared to the uniform linear arrays (ULAs) with the same number of physical elements.
	
	The prototype coprime array configuration, which consists of a pair of
	subarrays respectively equipped with $M$ and $N$ elements, where $M$ and $N$ are coprime integers, is able to resolve $\mathcal{O}(MN)$ sources with only $M+N-1$ physical elements \cite{2013Sparsity}. In order to
	fully utilize the DoFs offered by a coprime array, DoA estimation is implemented  using the virtual sensors of the difference coarray devised from the array data correlations
	\cite{Qin15}. The difference coarray derived from coprime arrays, however, usually contains multiple missing elements or ‘holes’, which lead to the model mismatch problem
	and degraded estimation performance \cite{Bou15}. A straightforward
	solution is to apply subspace-based spectral estimation algorithms, such as MUltiple SIgnal Classification
	(MUSIC), exploiting only the maximum contiguous segment of the
	difference coarray\cite{8008825}. However, such method underutilizes
	the array aperture and the available DoFs as a result of discarding the nonconsecutive virtual elements, thereby resulting in performance loss.
	
	The solutions of such problem can be basically divided into two types. One is to extend the maximum contiguous segment by coprime configuration design \cite{2017Generalized, 2020Two} or coprime array motion \cite{2020improved}. Nevertheless, these methods usually impose extra requirement on hardware complexity. The other proposes to fill the holes via interpolation, e.g., a gridless DoA estimation algorithm based on nuclear norm minimization (NNM) \cite{7539135, 2020Direct}.  Later, the positive semi-definite (PSD) structure of the covariance matrix \cite{2017Unified} is exploited to design the nuclear norm for covariance matrix construction.
	More recently, a virtual array interpolation algorithm based on the atomic norm minimization (ANM) \cite{Zhou18a} is presented to recover missing array data in a gridless manner.	
	
	In this letter, we propose a sparsity-aware algorithm for coprime array DoA estimation based on reconstruction of the covariance matrix using cyclic rank minimization from the measurements with missing holes. The proposed approach first interpolates the difference coarray and DoA estimation is reformulated as an ANM problem. Different from \cite{Zhou18a} which relaxes the NP-hard ANM problem into a convex one, we convert the ANM to an equivalent rank minimization problem of a Hermitian and Toeplitz matrix. While rank-minimization problems are common solved by relaxing the rank operation to a nuclear norm \cite{2020LowRank}, thus introducing approximation loss, we reformulate the rank-minimization problem by adopting a reformulation of the rank function \cite{8902019} which is multi-convex and equivalent to the original one.  The multi-convex optimization problem provides more accurate estimation of both the signal and noise subspaces for improved subspace-based DOA estimation. Numerical results verify that the proposed scheme achieves better estimation accuracy as compared to the state-of-the-art methods.
	
	Throughout this paper, we use lower-case (upper-case) bold characters to denote vectors (matrices), and vectors are by default in column orientation. Blackboard-bold characters denote standard sets of numbers and, in particular, $\mathbb{C}$ and $\mathbb{R}$ respectively denote the sets of complex and real numbers. $(\cdot)^{\rm T}$, $(\cdot)^*$, and $(\cdot)^{\rm H}$ respectively represent the transpose, complex conjugate, and conjugate transpose operators. ${\rm {tr}}(\cdot)$ denotes a trace operation of a matrix. ${\rm{diag}}\{\cdot\}$ represents a diagonal matrix that uses the entries of a vector as its diagonal entries. The vectorization operator ${\rm {vec}}(\cdot)$ sequentially stacks each column of a matrix. ${\bf{I}}$ denotes an identity matrix. Symbols $\circ$ and $\otimes$ respectively denote the Hardmard and Kronecker products. Symbol $\succeq$ represents PSD. ${\rm E}[\cdot]$ returns the expected value of a discrete random variable. $|\cdot|$ returns the cardinality of a set, and $|\cdot|_{\mathcal F}$ represents the Frobenius-norm. $\inf\{\cdot\}$ denotes the infimum of a given set.
	
	\section{Coprime Array Signal Model}
	
\begin{figure}[t]
		\centering
		\vspace{-2.2em}
		\includegraphics[width=0.78\linewidth]{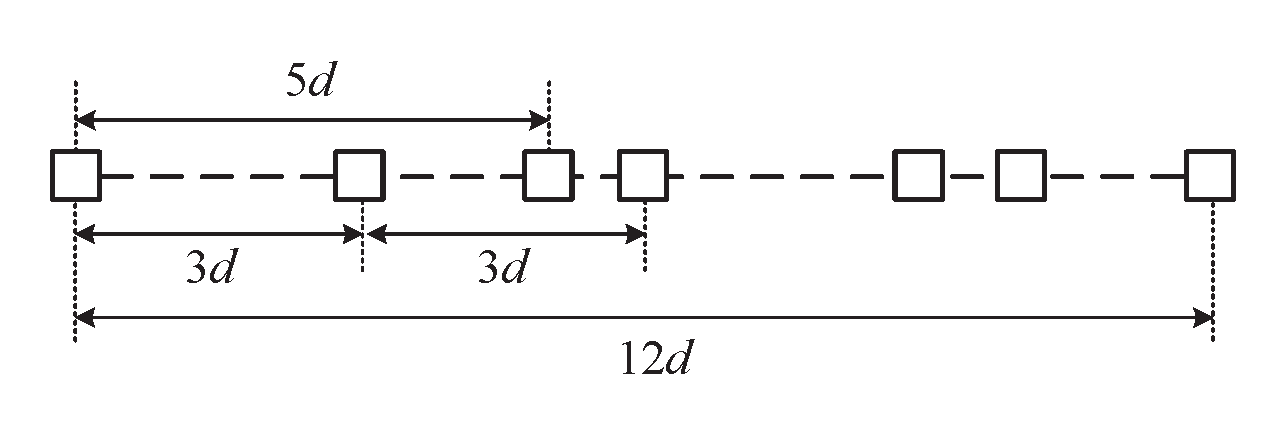}
		\vspace{-2.0em}
		\caption{Sensor geometry of a coprime array illustrated for $M$=3 and $N$=5.}
		\vspace{-0.5em}
		\label{F1}
	\end{figure}

	We consider a prototype coprime array structure shown in Fig.~\ref{F1}, where the sensor position are expressed as
	\begin{equation}\label{eq:1a}
	\mathbb{S}\!=\!\left\{Nmd,0\!\le\! m\!\le\! M-1\right\} \cup \left\{Mnd,0\!\le\! n\!\le\! N-1\right\},
	\end{equation}
	\hspace{-1.4em}
	where $M$ and $N$ are two coprime integers ($M\!<\!N$), $d=\lambda/2$ is the unit interelement spacing, and $\lambda$ denotes the wavelength. As such, the total number of physical sensors is $M\!+\!N\!-\!1$. We further assume that $K$ far-field, narrowband, and uncorrelated sources impinge from incident angles $\theta_k, \ k=1,\cdots,K$. The receive signal vector can be modelled as ${\bf{x}}(t)={\bf{A}}{\bf{s}}(t)+{\bf{n}}(t), \ t=1,\cdots,T$, where ${\bf{A}}=[{\bf{a}}(\theta_1),\cdots,{\bf{a}}(\theta_K)]\in\mathbb{C}^{(M+N-1)\times K}$ is the manifold matrix of the coprime array and ${\bf{a}}(\theta_k)=[1, \exp(-{\jmath} 2\pi d_2\sin{\theta_k}/\lambda), \ldots, \exp(-{\jmath} 2\pi d_{M+N-1}\sin{\theta_k}/\lambda)]^{\rm T}$ denotes the steering vector with $d_i\!\in\!\mathbb{S}$. In addition, ${\bf{s}}(t)\!=\![s_1(t),\ldots,s_K(t)]^{\rm T}$ with $s_k(t)$ denoting the waveform of the $k$-th signal. ${\bf{n}}(t)$ is an independent and identically distributed additive white Gaussian noise vector.
	
	The covariance matrix of the received signal vector can be written as ${\bf R}_{\bf x}={\rm E}[{\bf{x}}(t){\bf{x}}^{\rm H}(t)]={\bf{A}}{\bf R}_{\bf s}{\bf{A}}^{\rm H}+\sigma_n^2{\bf{I}}$, where ${\bf R}_{\bf s}$ represents the covariance matrix of the sources and $\sigma_n^2$ denotes the noise power. Note that ${\bf R}_ {\bf s}$ is a diagonal matrix, i.e., ${\bf R}_ {\bf s}={\rm{diag}}\{\sigma_1^2,\sigma_2^2,\cdots,\sigma_K^2\}$,  where $\sigma_k^2$ represents the power of the $k$-th source. In practice, the exact covariance matrix ${\bf R}_{\bf x}$ is unavailable and is estimated from the $T$ snapshots as
	\begin{equation}\label{eq:Rx_t}
	\hat{{\bf R}}_{\bf x}=\frac{1}{T}\sum\nolimits_{t=1}^{T}{\bf{x}}(t){\bf{x}}^{\rm H}(t).
	\end{equation}
	
	\section{Atomic norm of virtual array}
	
\begin{figure}[t]
		\centering
		\vspace{-1.0em}
		\includegraphics[width=0.98\linewidth]{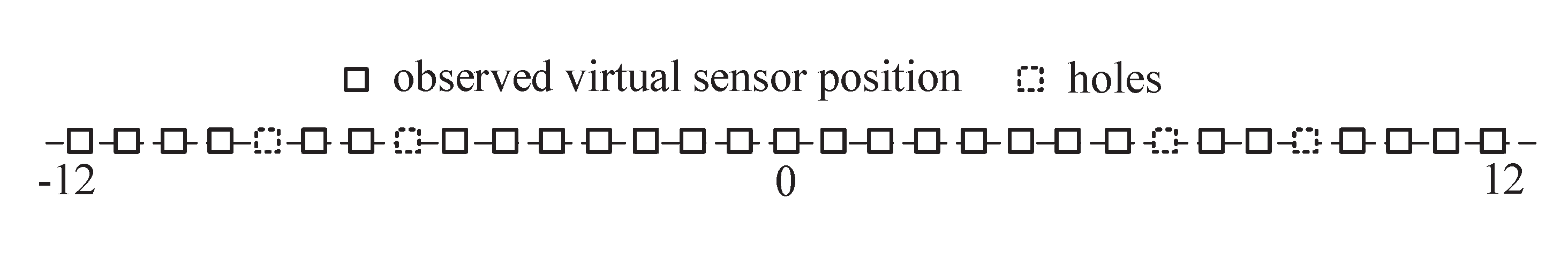}
		\vspace{-1.6em}
		\caption{Positions of the virtual sensors, $M$=3 and $N$=5.}
		\vspace{-0.5em}
		\label{F2}
	\end{figure}

	Vectorizing the covariance matrix ${\bf R}_{\bf x}$, the signal in the virtual sensor domain can be obtained as
	\begin{equation}
	{\bf v}={\rm{vec}}({{\bf R}_{\bf x}}) ={\bf A}_v{\bf p}+\sigma_n^2{\bf{i}},
	\end{equation}
	where ${\bf{A}}_{v}\!=\!{\bf{A}}\!\otimes\!{\bf{A}}^*$, ${\bf{p}}\!=\![\sigma_1^2,\sigma_2^2,\ldots,\sigma_K^2]^{\rm T}$, and ${\bf{i}}\!=\!\text{vec}\left({{\bf{I}}}\right)$.
	Each element of ${\bf v}$ corresponds to a virtual sensor whose position is determined by the difference between the physical sensor positions. The virtual sensor positions, illustrated in Fig.~\ref{F2}, can be derived by keeping all unique values from the difference set of the two coprime-integer sets,
	\begin{align}
	\mathbb{S}_{v}\subsetneq\mathbb{S}_{\rm {diff}}=\left\{s_d|s_d=\pm (Nm-Mn)d\right\}.
	\end{align}
	The output signal of the virtual sensor at position $s_d \in \mathbb{S}_v$ is computed as the cross-correlation between two physical sensors spaced apart by $s_d$. Thus, the equivalent virtual signal can be obtained by selecting the corresponding elements of ${\bf{v}}$ and removing the redundant ones as
	\begin{align} \label{barv}
	{\bf{\bar{v}}}={\bf{\bar{A}}}_{v}{\bf{p}}+\sigma_n^2{\bf{\bar{i}}},
	\end{align}
	where ${\bf{\bar{A}}}_{v}$ and ${\bf{\bar{i}}}$ are respectively sub-matrices of ${\bf{A}}_{v}$ and ${\bf{i}}$.
	
\begin{figure}[t]
		\vspace{-0.5em}
		\centering
		\includegraphics[width=0.8\linewidth]{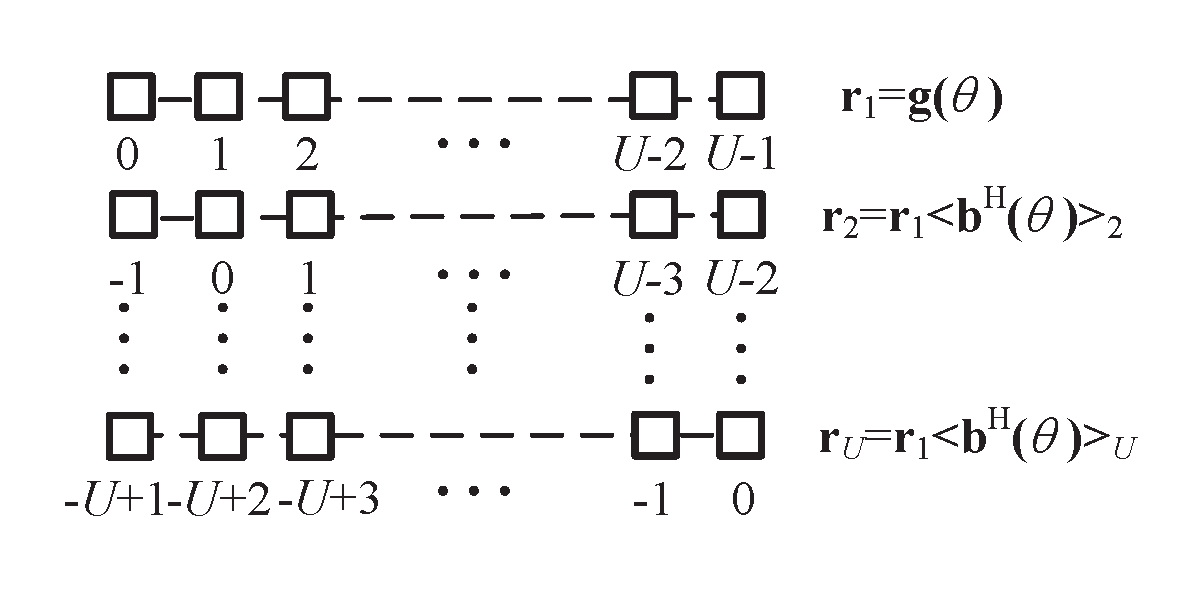}
		\vspace{-1.5em}
		\caption{Phase offsets among the virtual measurements of each sub-array, where $\left \langle{\bf{\cdot}}\right \rangle_{u}$ stands for the $u$-th element.}
		\vspace{-0.7em}
		\label{F3}
	\end{figure}

	In order to make the full use of the information contained in the non-uniform virtual array, we fill the holes in the difference coarray and obtain a virtual ULA via interpolation. We first initialize the interpolated virtual array signal ${\bf{v}}_{\rm I}$ as	
	\begin{align} \label{vi}
	\lbrack {\bf{v}}_{\rm I}\rbrack_i=
	\begin{cases}
	\quad \lbrack {\bf{\bar{v}}} \rbrack_i, & \quad i\in \mathbb{S}_{ {v}}, \\
	\quad 0, & \quad i\in \mathbb{S}_{\rm I}-\mathbb{S}_{ {v}},
	\end{cases}
	\end{align}
	where $\mathbb{S}_{\rm I}$ denotes the virtual ULA at all positions between $M(N-1)$ and $-M(N-1)$, and $[\cdot]_{i}$ represents the virtual sensor at position $id$.
	%
	Then, based on the idea of atomic norm of multiple virtual measurements \cite{Zhou18a}, the interpolated virtual array $\mathbb{S}_{\rm I}$ is divided into $U\!=\!(|\mathbb{S}_{\rm I}|\!+\!1)/2\!=\!M(N-1)\!+\!1$ overlapping sub-arrays, each with $U$ contiguous virtual sensors, as shown in Fig.\;\ref{F3}. Accordingly, the virtual signal vector ${\bf{v}_{\rm I}}$ of the interpolated virtual array $\mathbb{S}_{\rm I}$ is divided into $U$ sub-vectors $\left\{{\bf{r}}_1,\ldots,{\bf{r}}_U\right\}$ to form Hermitian and Toeplitz matrix  ${\bf{V}}=[{\bf{r}}_1,\ldots,{\bf{r}}_U]\in\mathbb{C}^{U\times U}$.
	
	An atom that represents ${\bf{V}}$ can be expressed as ${\bf{G}(\theta)}={\bf{g}(\theta)}{\bf{b}^{\rm H}(\theta)}\in\mathbb{C}^{U\times U}$ for $\theta\in[-90^{\circ},90^{\circ}]$, where ${\bf{g}(\theta)}$ denotes the steering vector of the first sub-array of $\mathbb{S}_{\rm I}$ and is referred to as the reference virtual array. In addition, ${\bf{b}(\theta)}=[1,\exp(-\jmath\pi \sin{\theta}),\ldots,\exp(-\jmath\pi(U-1)\sin{\theta})]^{\rm T}$ represents the phase offsets between $U$ sub-arrays. Thus, the corresponding atom set is given as
	\begin{align}
	\mathbb{A}=\left\{{\bf{G}(\theta)}|\theta\in[-90^{\circ},90^{\circ}]\right\}.
	\end{align}
	The smallest number of atoms for representing the virtual measurements ${\bf{V}}$ can be defined as
	\begin{align} \label{atom1}
	||{\bf{V}}||_{\mathbb{A},0}=\inf \limits_{K} \left\{{\bf{V}}=\sum\nolimits_{k=1}^{K}{p_{k}}{{\bf{G}}(\theta_{k})},p_{k}\ge 0 \right\}.
	\end{align}
	
	\section{DoA Estimation via Toeplitz Matrix Reconstruction}
	
	In this section, we develop a novel DoA estimation algorithm based on the cyclic low-rank recovery of the virtual array covariance matrix which is Hermitian and Toeplitz. The proposed algorithm converts the ANM problem into a dual-variable cyclic rank minimization problem with a partial closed-form solution. Once the covariance matrix, which corresponds to a ULA, is recovered, the source DoAs can be readily estimated by using subspace-based DoA estimation algorithms such as MUSIC. To formulate the cyclic rank minimization problem, we first derive the following theorem.
	
	$\textit{Theorem 1}$: Let ${\bf{T({\bf{z}}}})\in\mathbb{C}^{U\times U}$ be a Hermitian and Topelitz covariance matrix of the signals received by a virtual ULA, with vector ${\bf{z}}$ as the first column of ${\bf T}({\bf z})$.	Define another Hermitian and Topelitz matrix ${\bf{M}}$. Then, problem \eqref{atom1} is equivalent to
	\begin{align}    \label{first}
	\min \limits_{{\bf{z}},{\bf{M}}} \qquad &{{\rm{rank}}[{\bf{T({\bf{z}})}}]} \nonumber \\
	\text{subject to} \quad &
	\begin{bmatrix}
	{\bf{T({\bf{z}}}}) & {\bf{V}}    \\
	{\bf{V}}^{\rm H} & {\bf{M}}
	\end{bmatrix}
	\succeq 0.
	\end{align}
	\begin{IEEEproof}
		Denote by $r_{\text{opt}}$ the minimum rank obtained from \eqref{first}. We first show $r_{\text{opt}}\le||{\bf{V}}||_{\mathbb{A},0}$. Denote $||{\bf{V}}||_{\mathbb{A},0}=m$ and assume that the decomposition ${\mathbf{V}}=\sum_{k=1}^{m}p_k\mathbf{g}\left(\theta_{k}\right)\mathbf{b}^{\rm H}\left(\theta_{k}\right)$ with $p_k>0$ achieves $||{\bf{V}}||_{\mathbb{A},0}$. We further assume that $\mathbf{z}=\sum_{k=1}^{m}p_k\mathbf{g}\left(\theta_{k}\right)$ and $\mathbf{M}=\sum_{k=1}^{m}p_k\mathbf{b}\left(\theta_{k}\right)\mathbf{b}^{\rm H}\left(\theta_{k}\right)$, such that $\mathbf{T}\left(\mathbf{z}\right)=\sum_{k=1}^{m}p_k\mathbf{g}\left(\theta_{k}\right)\mathbf{g}^{\rm H}\left(\theta_{k}\right)\succeq 0$. Then, the matrix in the constraint of \eqref{first} can be expressed as
		\begin{align}\label{subproof1}
		\mathbf{J}\triangleq
		\begin{bmatrix}
		{\bf{T({\bf{z}}}}) \!&\! {\bf{V}}    \\
		{\bf{V}}^{\rm H} \!&\! {\bf{M}}
		\end{bmatrix}
		\!=\!\sum_{k=1}^{m}p_k\!
		\begin{bmatrix}
		\mathbf{g}\left(\theta_k\right)   \\
		\mathbf{b}\left(\theta_k\right)
		\end{bmatrix}
		\begin{bmatrix}
		\mathbf{g}^{\rm H}\left(\theta_k\right)\!&\!\mathbf{b}^{\rm H}\left(\theta_k\right)
		\end{bmatrix}
		\succeq 0.
		\end{align}
		This implies that $\mathbf{T}\left(\mathbf{z}\right)$ is a sub-matrix of matrix $\mathbf{J}$ in the constraint which can be written as an $m$-fold factorization. As such, its rank satisfies $r_{\text{opt}}\le m=||{\bf{V}}||_{\mathbb{A},0}$.
		
		On the other hand, suppose that the optimal solutions of \eqref{first} are $\mathbf{z}_\text{opt}$ and $\mathbf{W}_\text{opt}$. If $\mathbf{T}\left(\mathbf{z_\text{opt}}\right)=\mathbf{D}\mathbf{C}\mathbf{D}^{\rm H}$ is a Vandermonde decomposition, the positive semidefiniteness
		of matrix $\mathbf{J}$ implies that $\mathbf{V}$ is in the range of $\mathbf{D}$. This in turn reveals that $\mathbf{V}$ can be expressed as a combination of at most $r_\text{opt}$ atoms \cite{6576276}, i.e.,  $r_{\text{opt}}\ge||{\bf{V}}||_{\mathbb{A},0}$. The proof is finished.
	\end{IEEEproof}
	
	By noting the equivalence between the recovered $\mathbf{V}$ and $\mathbf{z}$, as shown in \cite{Zhou18a}, (\ref{first}) can be further expressed as
	\vspace{-0.5em}
	\begin{align} \label{rank1}
	\min \limits_{{\bf{z}}} \qquad &{{\rm{rank}}[{\bf{T({\bf{z}})}}]} \nonumber \\
	\text{subject to} \quad &
	\begin{bmatrix}
	{\bf{T({\bf{z}}}}) & {\bf{z}}    \\
	{\bf{z}}^{\rm H} & U^{-1}{\rm{tr}}[{\bf{T({\bf{z}}}})]
	\end{bmatrix}
	\succeq 0.
	\end{align}
	To solve the rank-minimization problem \eqref{rank1} and prevent the approximation loss, we adopt a reformulation of the rank function\cite{8902019} which is equivalent to the original one. Let $\gamma \textgreater 0$ be a positive constant, $\mathbf{W}\succeq 0$ be a PSD matrix, and define function $\text{f}[{\bf{W}},{\bf{T}}(\bf{z}),\gamma]$ as
	\begin{align}
	\text{f}[{\bf{W}},{\bf{T}}({\bf{z}}),\gamma]={\gamma}^{-2}(\Vert{\bf{W}-\gamma{\bf{I}}}\Vert)^{2}_{\mathcal F}+2{\rm{tr}}[{\bf{WT}({\bf{z}})}].
	\end{align}
	Then, the rank-minimization problem is equivalent to minimizing  $\text{f}[{\bf{W}},{\bf{T}}({\bf{z}}),\gamma]$ under the constraints  ${\rm{tr}}[{\bf{WT}}({\bf{z}})]\leq 0$ and ${\bf{W}}\succeq 0$. We further combine the initialized signal in \eqref{vi} as the reference. Then, \eqref{rank1} is reformulated as
	\vspace{-0.5em}
	\begin{align} \label{g}
	\min \limits_{{\bf{z}},{\bf{W}}} \qquad &\text{f}[{\bf{W}},{\bf{T}}({\bf{z}}),\gamma] \nonumber \\
	\text{subject to} \quad &
	{\Vert{\bf{T}}({\bf{z}})\circ{\bf{B}}-\tilde{\bf{R}}_{v}\Vert}_{\mathcal F}\le\eta ,\nonumber\\
	& {\rm{tr}}[{\bf{WT}}({\bf{z}})]\leq 0, {\bf{W}}\succeq 0,   \nonumber \\
	&	
	\begin{bmatrix}
	{\bf{T({\bf{z}}}}) & {\bf{z}}    \\
	{\bf{z}}^{\rm H} & U^{-1}{\rm{tr}}[{\bf{T({\bf{z}}}})]
	\end{bmatrix}
	\succeq 0,
	\end{align}
	where $\tilde{\bf{R}}_{v}={\bf{T}}({\bf{r}}_1)$, which can be calculated from the first reference sub-array of $\mathbf{v}_\text{I}$, is formulated as the reference virtual array covariance matrix of the initialized signal, and ${\bf{B}}\in\mathbb{C}^{U\times U}$ is a binary matrix that is used to distinguish the zero (interpolated) and non-zero (derived) statistics in $\tilde{\bf{R}}_{v}$ after the initial virtual array interpolation. By observing the results in \eqref{g}, the following remarks are in order.
	
	$\textit{Remark 1}$:
	Note that function $\text{f}[{\bf{W}},{\bf{T}}({\bf{z}}),\gamma]$ is multi-convex and, thus, (\ref{g}) is a multi-convex optimization problem. That is, when ${\bf{z}}$ or ${\bf{W}}$ is fixed, ({\ref{g}}) becomes a convex function of ${\bf{W}}$ or ${\bf{z}}$, respectively. Hence, we can alternatively optimize variables ${\bf{z}}$ and ${\bf{W}}$ by fixing one when updating the other.
	
	$\textit{Remark 2}$:
	Optimization problem \eqref{g} introduces a new variable $\mathbf{W}$. As both $\mathbf{T}\left(\mathbf{z}\right)$ and $\mathbf{W}$ are PSD, $\text{tr}\left[\mathbf{W}\mathbf{T}\left(\mathbf{z}\right)\right]\le 0$ implies that $\mathbf{W}$ must be in the null-space of $\mathbf{T}\left(\mathbf{z}\right)$. That means, the optimization of both the signal subspace and null-space (which can also be considered as noise subspace in the area of DoA estimation) are taken into account in \eqref{g}.
	
	$\textit{Remark 3}$:
	Since the minimization of $\text{f}[{\bf{W}},{\bf{T}}({\bf{z}}),\gamma]$ involves the minimization of ${\rm{tr}}[{\bf{W}}{\bf{T({\bf{z}}}})]$, the latter can be utilized as a stopping condition for the algorithm.
	
	We further reformulate the first constraint in \eqref{g} as a regularization term of the optimization function controlled by $\mu$. As such, the proposed alternative optimization problems for the cyclic minimization algorithm \cite{8902019} to minimize as per (\ref{g}) are respectively given as
	\begin{align}   \label{iter1}
	{\bf{z}}_{(i)}=&\mathop{\arg\min_{{\bf{z}}}} \text{f}[{\bf{W}}_{(i-1)},{\bf{T}}({\bf{z}}),\gamma]+\mu{\Vert{\bf{T}}({\bf{z}})\circ{\bf{B}}-\tilde{\bf{R}}_{v}\Vert}_{\mathcal F} \nonumber \\
	&\text{subject to} \quad
	\begin{bmatrix}
	{\bf T}({\bf z}) & {\bf{z}}    \\
	{\bf{z}}^{\rm H} & U^{-1}{\rm{tr}}[{\bf T}({\bf z})]
	\end{bmatrix}
	\succeq 0,
	\end{align}
	and
	\begin{align}   \label{iter2}
	&{\bf{W}}_{(i)}=\mathop{\arg\min_{{\bf{W}}}} \text{f}[{\bf{W}},{\bf{T}}({\bf{z}}_{(i)}),\gamma]+\mu{\Vert{\bf{T}}({{\bf{z}}_{(i)}})\circ{\bf{B}}-\tilde{\bf{R}}_{v}\Vert}_{\mathcal F} \nonumber \\
	& \quad \quad \quad \quad \quad \quad \quad
	\text{subject to} \quad
	{\bf{W}}\succeq 0,
	\end{align}
	where ${\bf{z}}_{(i)}$ and ${\bf{W}}_{(i)}$ denote their values in the $i$-th iteration, and the algorithm terminates when ${\rm{tr}}[{\bf{W}}_{(i)}{\bf{T}}({\bf{z}}_{(i)})]$ converges or when the maximum number of iterations $N^{\text{iter}}_{\rm{max}}$ is reached. In addition, the two iterative processes ({\ref{iter1}}) and ({\ref{iter2}}) respectively correspond to the optimization of the signal subspace and null-space, since ${\bf{z}}$ is in the range of ${\bf{T}}({\bf{z}})$ while ${\bf{W}}$ is in the null-space of ${\bf{T}}({\bf{z}})$.
	
	Both (\ref{iter1}) and (\ref{iter2}) are convex SDP problems. However, we can further reduce the complexity because a closed-form of (\ref{iter2}) can be obtained by performing the eigen-decomposition of the Hermitian and Toeplitz matrix ${\bf{T(\bf z)}}\in\mathbb{C}^{U\times U}$ as
	\begin{align}
	{\bf{T(\bf z)}}={\bf{U}}{\boldsymbol \Sigma}{\bf{U}}^{\rm H}, \quad {\boldsymbol \Sigma}={\rm{diag}}[\left\{\lambda_{u}[{\bf{T(\bf z)}}]\right\}^U_{u=1}],
	\end{align}	
	where ${\bf{U}}\in\mathbb{C}^{U\times U}$ is a unitary matrix and $\lambda_{1}[{\bf{T(\bf z)}}]\!\ge\!\lambda_{2}[{\bf{T(\bf z)}}]\!\ge\cdots\!\ge\!\lambda_{U}[{\bf{T(\bf z)}}]\!\in\!\mathbb{R}$ are positive eigenvalues. Taking into account the similarities among the solutions of such problems \cite{8902019, 2010A}, the global optimal solution of problem ({\ref{iter2}}) can be derived as
	\begin{equation}\label{eq:opt_W}
	{\bf{W}}^{\text{opt}}_{(i)}={\Omega}_{\gamma,0}[{\bf{T}}[{\bf{z}}_{(i)}]],
	\end{equation}
	where the operator ${\Omega}_{\gamma,0}[{\bf{T(\bf z)}}]$ is defined as
	\begin{align}
	\!{\Omega}_{\gamma,0}[{\bf{T(\bf z)}}]\triangleq{\bf{U}}\boldsymbol{\Gamma}_{{\boldsymbol\Sigma}}{\bf{U}}^{\rm H},
	\end{align}
	and $\boldsymbol{\Gamma}_{{\boldsymbol\Sigma}}\!=\!{\rm{diag}}[\{\text{max} [\gamma-\lambda_{U-u+1}[{\mathbf{T}(\mathbf{z})}],0]\}^U_{u=1}]$.
	
	As for \eqref{iter1}, there is no closed-form solution but its can be solved efficiently by using available SDP solvers (e.g., SDPT3, SeDuMi). As a result, the minimization problem ({\ref{g}}) can be efficiently solved within a few iterations. After the desired covariance matrix ${\bf{T}}({\bf{z}})$ is recovered, we can estimate the DoAs by applying MUSIC based on ${\bf{T}}({\bf{z}})$. The proposed method is summarized in {\bf Algorithm~\ref{alg}}.	
	\begin{algorithm}[h]
		{\small
			\caption{DoA estimation algorithm based on rank minimization-based Toeplitz covariance matrix reconstruction.}
			\label{alg}
			\textbf{Input:} Receive signal $\left\{{\bf x}(t)\right\}^{T}_{t=1}.$\\
			\textbf{Output:} DoAs $\theta_{k},k=1,...,K.$ \\
			\textbf{Initialize:} ${\bf W} \leftarrow$ A random Hermitian matrix, and define $\gamma$, $\mu$, $\epsilon$, and $N^{\text{iter}}_{\rm{max}}$.
			\begin{algorithmic}[1]
				\item Derive the covariance matrix ${\bf R}_x$ using \eqref{eq:Rx_t};
				\item Obtain the equivalent virtual signal ${\bf \bar{v}}$ using (\ref{barv});
				\item Initialize the interpolated virtual array signal ${\bf v}_{\rm I}$ using (\ref{vi});
				\item Construct the reference virtual covariance matrix as $\tilde{\bf{R}}_{v}={\bf{T}}({\bf{r}}_{1})$;
				\item Use a binary matrix ${\bf B}$ to distinguish the elements in the reference virtual array;
				\For{i=1 to $N^{\text{iter}}_{\rm{max}}$}
				\State Solve the rank minimization problem (\ref{iter1}) and yield ${\bf{T}}({\bf{z}}_{(i)})$;
				\While{$\lvert{\rm{tr}}[{\bf{W}}_{(i)}{\bf{T}}({\bf{z}}_{(i)})]-{\rm{tr}}[{\bf{W}}_{(i-1)}{\bf{T}}({\bf{z}}_{(i)})]\rvert > \epsilon$} \\
				\qquad\qquad Optimize ${\bf W}_{(i)}$ using \eqref{eq:opt_W};
				\EndWhile
				\EndFor
				\item Perform MUSIC to estimate the DoAs $\theta_{k}$.
		\end{algorithmic}}
	\end{algorithm}
	
	\vspace{-1em}
	\section{Simulation Results}
	
	In this section, the effectiveness of the proposed scheme is demonstrated through numerical simulations. Let $M=3$ and $N=5$ be the coprime integer pair, and a total number of $M+N-1=7$ physical sensors are located at $\{0,3d,5d,6d,9d,10d,12d\}$. We set the parameters as $\gamma=0.05$, $\mu=40$, $\epsilon=10^{-4}$, and $N^{\text{iter}}_{\rm{max}}=50$.
	
\begin{figure}[t]
		\centering	
		\vspace{-0.5em}
		\subfloat[]{\includegraphics[width=0.5\linewidth]{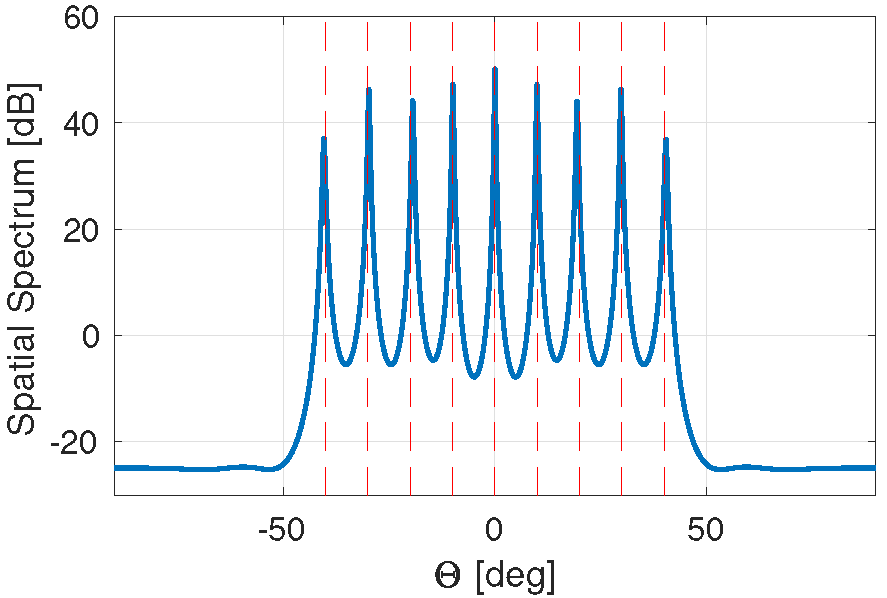}}
		\subfloat[]{\includegraphics[width=0.5\linewidth]{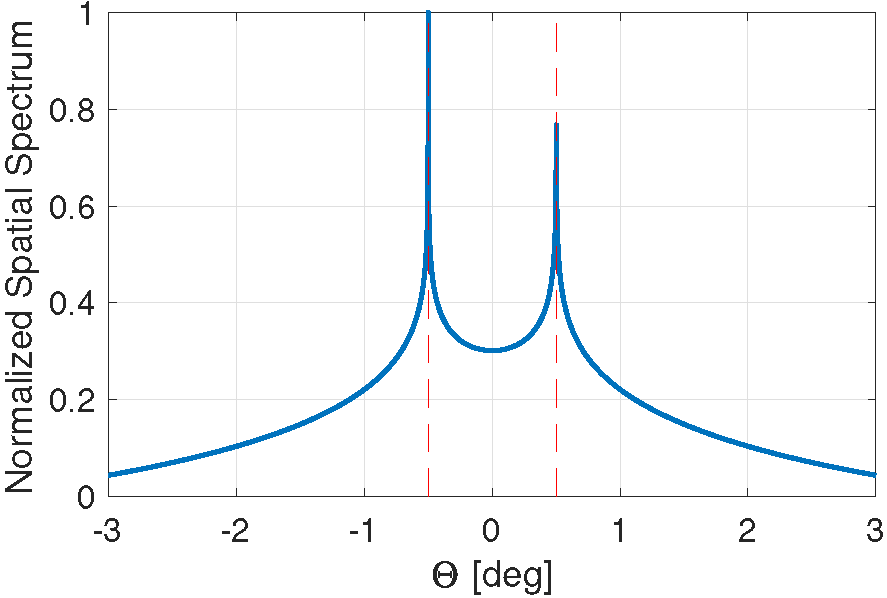}}
		\vspace{-0.5em}
		\caption{Spatial spectrum obtained by the proposed algorithm. (a) DoF capability. (b) Resolution capability.}
		\vspace{-1.0em}
		\label{SpatSpec}
	\end{figure}

	In the first example, we confirm the high number of DoFs achieved by the proposed algorithm. Assume nine uncorrelated equal-power incident sources that are uniformly distributed in $[-40^{\circ},40^{\circ}]$ with $30\;{\rm{dB}}$ signal-to-noise ratio (SNR), and $500$ snapshots are acquired. The vertical dashed red lines denote the true DoAs of the incident sources. It is observed in Fig.~\ref{SpatSpec}(a) that the proposed scheme is able to accurately estimate the DoAs of all nine sources with only seven physical sensors.
	
	Next, we examine the resolution performance of the proposed method by considering two closely separated uncorrelated sources impinging from directions $\theta_{1}=-0.5^{\circ}$ and $\theta_{2}=0.5^{\circ}$. The input SNR and the number of snapshots remain unchanged. As shown in Fig.~\ref{SpatSpec}(b), the proposed algorithm resolves both sources in their true directions with shape peaks.
	
\begin{figure*}[!htpb]
		\centering
		\subfloat[]{\includegraphics[width=0.39\linewidth]{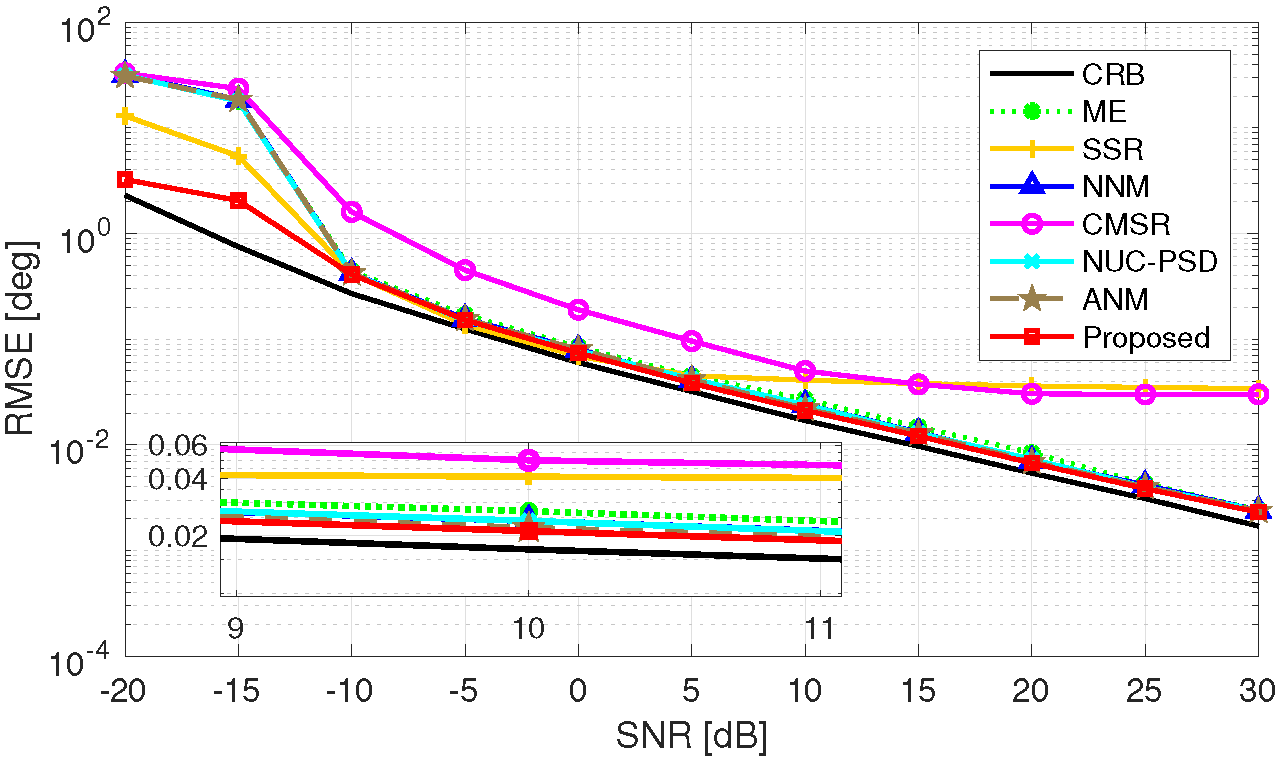}}
		\hfil
		\subfloat[]{\includegraphics[width=0.39\linewidth]{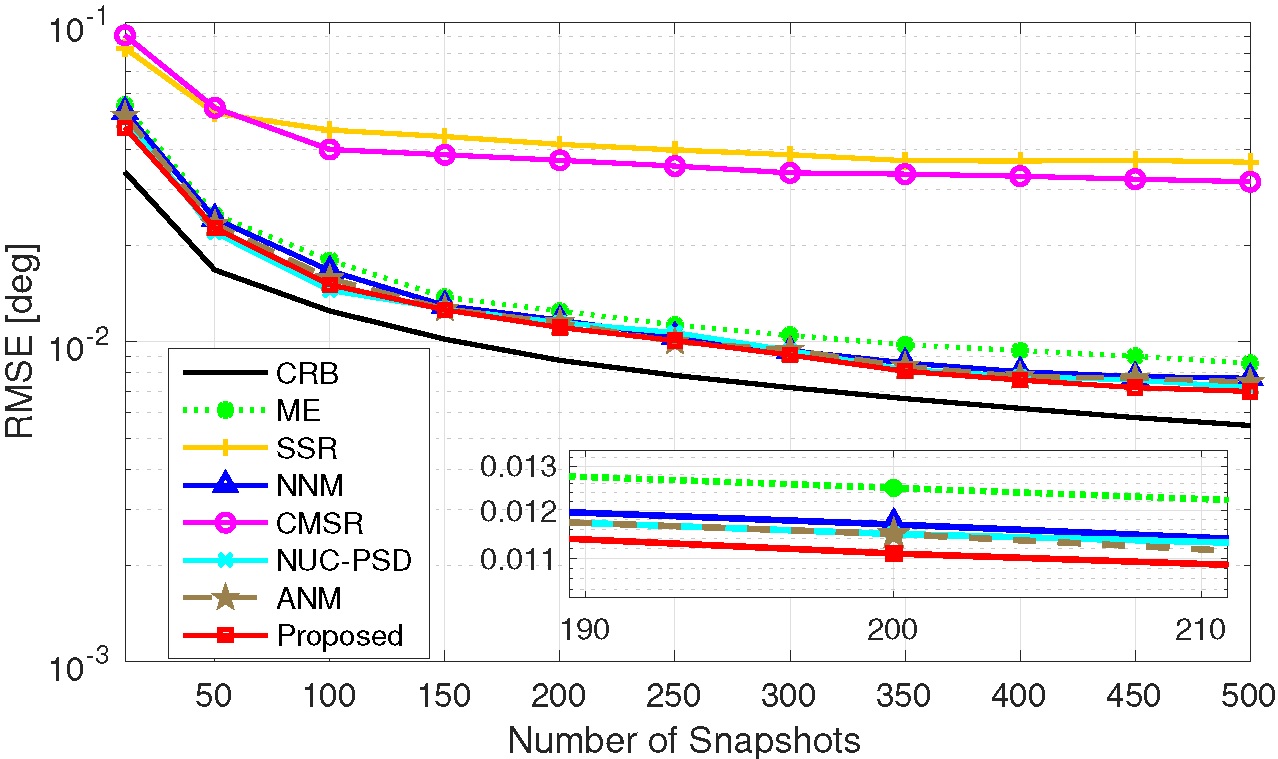}}\\
		\vspace{-1.2em}
		\subfloat[]{\includegraphics[width=0.39\linewidth]{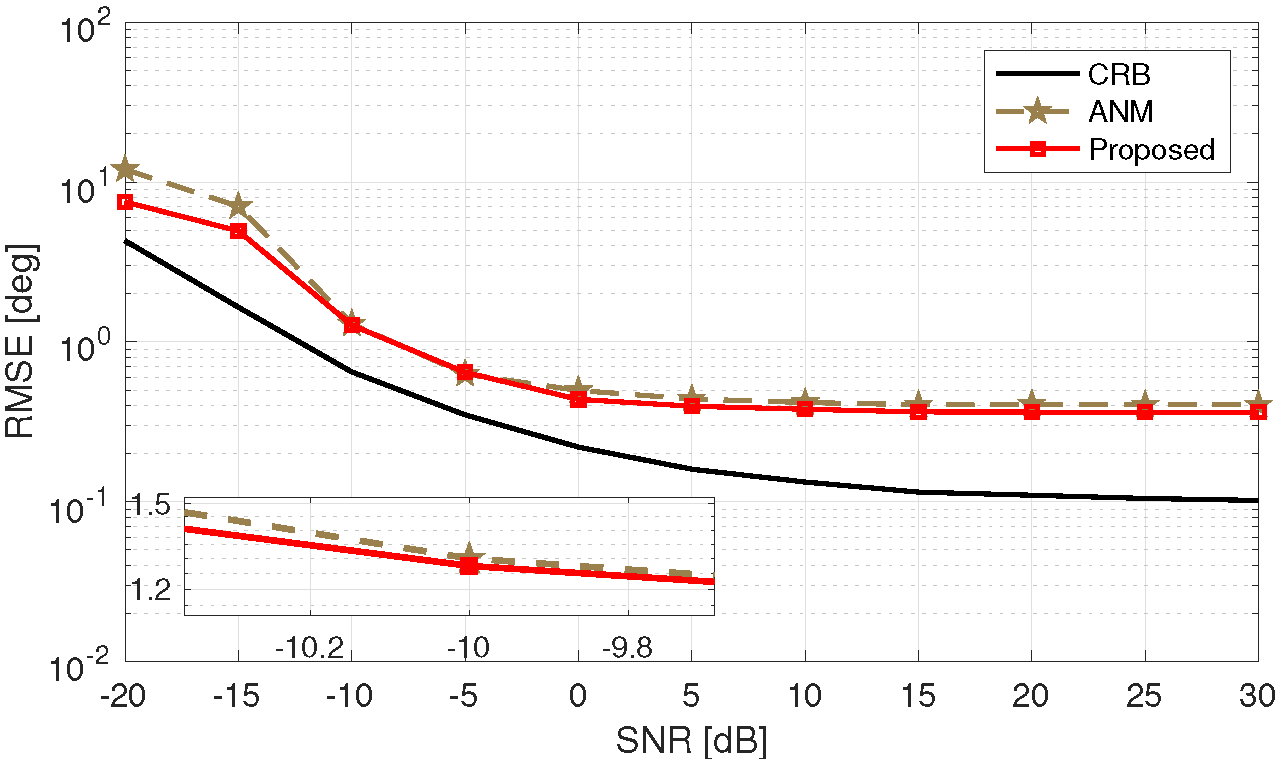}}
		\hfil
		\subfloat[]{\includegraphics[width=0.39\linewidth]{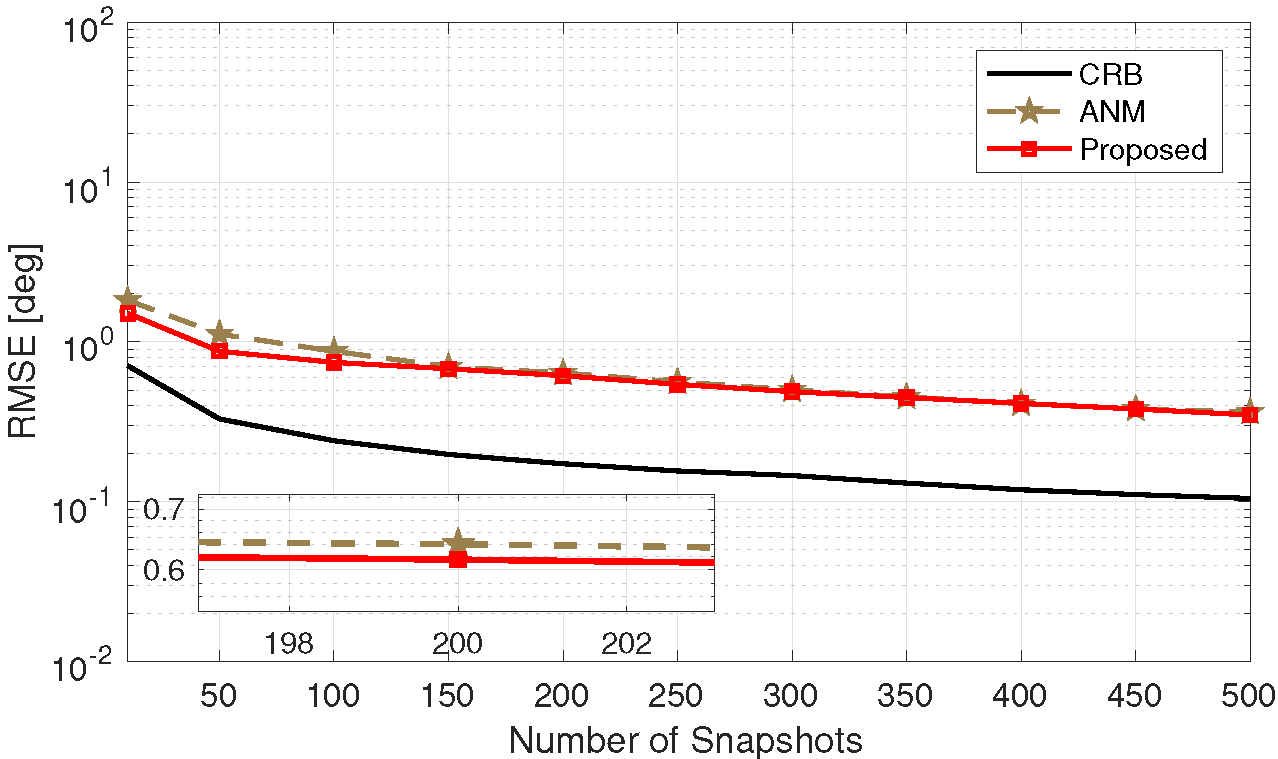}}
		\vspace{-0.3em}
		\caption{Performance comparisons of different methods. (a) RMSE versus SNR, single source; (b) RMSE versus number of snapshots, single source; (c) RMSE versus SNR, nine sources; (d) RMSE versus number of snapshots, nine sources.}
		\vspace{-1em}
		\label{fig:stat}
	\end{figure*}

	In the following, the root mean square error (RMSE) of the estimated DoAs obtained from the proposed algorithm is compared with the Cram\'{e}r-Rao bound (CRB) \cite{2002MUSIC} and those achieved by the state-of-the-art DoA estimation algorithms, including the sparse signal reconstruction (SSR) algorithm \cite{2013Sparsity}, the NNM algorithm \cite{7539135}, the NNM with PSD constraint (NUC-PSD) algorithm \cite{2017Unified}, the maximum entropy (ME) algorithm \cite{2017Unified}, the ANM algorithm \cite{Zhou18a}, and the covariance matrix sparse reconstruction (CMSR) algorithm \cite{2017Source}. The direction of the incident signal is randomly generated from the Gaussian distribution $\mathcal{N}(0^\circ,(1^\circ)^2)$, and the results are computed using $1,000$ Monte Carlo trials. We first fix the number of snapshots to $500$ and let the input SNR vary between $-20\;{\rm{dB}}$ and $30\;{\rm{dB}}$. As indicated in Fig.~\ref{fig:stat}(a), compared with the competitive algorithms in the case of low SNR which varies between $-20\;{\rm{dB}}$ and $-10\;{\rm{dB}}$, the RMSE of the proposed algorithm is significantly lower and much closer to the CRB. As the SNR increases, the RMSE of all the algorithms, except SSR and CMSR, gradually decreases, but the proposed method provides the lowest RMSE results. The floor of the RMSE performance observed for SSR and CMSR is because these methods are grid-based algorithms and thus suffer from performance loss due to the basis mismatch problem. When we fix the input SNR to 20 dB and vary the number of snapshots, the results depicted in Fig.~\ref{fig:stat}(b) confirm again that SSR and CMSR render high errors, whereas all other methods achieve similar RMSE performance with the proposed algorithm slightly outperforming others.
	
	Next, we consider the scenario where the number of sources is greater than the number of physical sensors. The parameters of the sources are identical to those of Fig.~\ref{SpatSpec}(a). In this, some methods do not resolve all sources \cite{Zhou18a}. We compare the RMSE results of the proposed method with that of the ANM and the coarray CRB in Figs.~\ref{fig:stat}(c) and \ref{fig:stat}(d), respectively with respect to the input SNR and the number of snapshots.  It is observed in both Fig.~\ref{fig:stat}(c) and Fig.~\ref{fig:stat}(d) that the RMSE of the proposed method is consistently lower than that obtained by the ANM with a small margin.
	
	Regarding the computational complexity, simulation results show that proposed method usually converges in 3 iterations. The ANM takes 35 seconds to compute 50 Monte Carlo trials on a 16 GB Intel(R) Core(TM) i7-4980 HQ CPU, while the proposed method takes 114 seconds. Compared to the ANM which only optimizes the signal subspace, the computational complexity of the proposed method is slightly higher as it needs to optimize both the signal and noise subspaces.
	
	\section{Conclusion}
	
	In this paper, we presented a novel DoA estimation algorithm for coprime array. To fully utilize the underlying received information in the presence of missing elements in the difference coarray, interpolation is performed and a dual-variable rank minimization problem is formulated. We recast the problem as a multi-convex form and developed an alternative optimization mechanism to solve the problem through cyclic iterations. Improved DoA estimation performance based on the reconstructed Toeplitz covariance matrix was confirmed by numerical simulations.

\end{document}